\documentclass[twocolumn,prl,showpacs]{revtex4}
\usepackage{psfrag}
\usepackage{graphicx}
\usepackage{amsmath}
\usepackage{amssymb}

\begin{document}
\title{Interference and $k$-point sampling in
  the supercell approach to \\phase-coherent transport.}

\author{K. S. Thygesen}
\affiliation{Center for Atomic-scale Materials Physics, \\
  Department of Physics, Technical University of Denmark, DK - 2800
  Kgs. Lyngby, Denmark}
\author{K. W. Jacobsen}
\affiliation{Center for Atomic-scale Materials Physics, \\
  Department of Physics, Technical University of Denmark, DK - 2800
  Kgs. Lyngby, Denmark}
\date{\today}

\begin{abstract}
We present a systematic study of interference and $k$-point sampling
effects in the supercell approach to phase-coherent electron
transport. We use a representative tight-binding model to show that interference
between the repeated images is a small effect compared to the error
introduced by using only the $\Gamma$-point for a supercell
containing (3,3) sites in the transverse plane. An insufficient
$k$-point sampling can introduce strong but unphysical features in the
transmission function which can be traced to the presence of van Hove
singularities in the lead.
We present a first-principles calculation of the transmission through a
Pt contact which shows that the $k$-point sampling is also important for
realistic systems.

\end{abstract}

 \maketitle 
The growing interest for
exploring the electronic conduction properties of nano-structures has
led to the development of a variety of numerical methods aimed at
describing phase-coherent electron transport in atomic-scale systems from
first principles.  Many of these methods are based on a supercell
approach where the system of interest is defined inside a finite
computational cell which is then repeated periodically in the
directions perpendicular to the transport direction while the
appropriate open boundary conditions needed to describe the coupling
to the electrodes are imposed in the parallel
direction~\cite{korn02,jelinek03,thygesen03,calzolari04,choi99,nielsen03,khomyakov04,taylor03,brandbyge02,tsukamoto02,stokbro03}.
Within the supercell approach, one is obviously limited to consider
transport through an infinite array of contacts, however, in practice
this limitation is usually ignored. Whether the supercell approach can
provide a good description of transport through a single, isolated
contact depends on the degree of interference between the repeated
images.  As the transverse dimensions of the supercell are increased
the interference is expected to decrease and the result should
approach that of a single contact. For practical first principles
calculations this limit is very hard to realize due to the
computational cost associated with an enlargement of the supercell and
therefore relatively small supercells containing e.g. $3\times3$ atoms
in the transverse plane are typically used as a
compromise~\cite{jelinek03,thygesen03,nielsen03,khomyakov04,taylor03,brandbyge02,stokbro03}.

A second approximation which is also commonly used, is to evaluate
the transmission function only at the
$\Gamma$-point of the two-dimensional Brillouin zone (BZ) of the plane
perpendicular to the transport direction. For
example this approximation has been applied in transport studies of
Au~\cite{brandbyge02} and Na~\cite{tsukamoto02} wires and to
molecular contacts containing di-thiol benzene~\cite{stokbro03}.
Again, when the transverse dimensions of the supercell are
increased the $\Gamma$-point approximation eventually becomes valid. 
While the role of $\bold k$-point sampling for total energy
calculations is well understood, its influence on the transmission function has, to
our knowledge, so far not been systematically explored. In a simple picture
the transmission function is similar to a density of states
(DOS) and as such it should be more sensitive to the $\bold k$-point sampling than
the total energy which is an integral of the DOS.

In this paper we investigate the effects of $\bold k$-point sampling
and interference on the transmission function by means of two
examples.  The first example is based on a tight-binding (TB) model.
The simplicity of this model allows for a complete separation of
interference and $\bold k$-point sampling effects and thus provides an
opportunity to study the relative importance of these factors for the
transmission. Our results show that even for a small cell containing
$3\times 3$ sites in the transverse directions, interference effects
are not very significant compared to the error introduced by an
insufficient $\bold k$-point sampling. The second example is a first principles calculation for
an atomic Pt contact demonstrating that $\bold k$-point sampling is
essential also for realistic systems.
\begin{figure}[!b]
\includegraphics[width=0.68\linewidth]{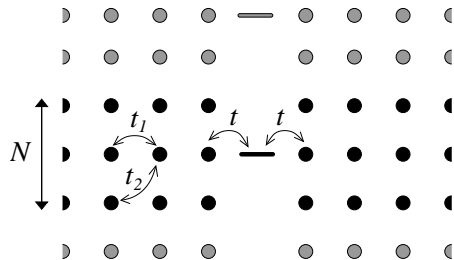}
\caption[cap.wavefct]{\label{fig1} Two-dimensional illustration
  of the TB model. A supercell is indicated by the
  black dots. The molecular states have on-site energy $\varepsilon_a$
  and are coupled symmetrically to semi-infinite cubic
  lattices via the hopping parameter $t$. All on-site
  energies of the lattice are zero and the hopping connecting $i$th nearest neighbors is denoted $t_i$.}
\end{figure}

We consider a tight-binding model for resonant transport through an
array of identical states which we refer to as "molecular states".
The leads are described by semi-infinite cubic lattices with zero on-site energies  and
hopping parameter $t_i$ between $i$th nearest neighbors. In all
calculations we use the parameters $t_1=1.0$, $t_2=0.4$, $t_3=0.1$ to
describe the leads. The molecular
states have on-site energy $\varepsilon_a$ and are coupled to the
leads via a single lattice site by the hopping parameter $t$.
The number of lattice sites between the periodically repeated molecular
states is $N$ in each direction. A two-dimensional version of the model is illustrated in Fig.~\ref{fig1}.

The transmission function is calculated from the general formula~\cite{meirwingreen} 
\begin{equation}\label{eq.trans}
T(\varepsilon)=\text{Tr}\big [G^r_C(\varepsilon) \Gamma_L(\varepsilon) G^a_C(\varepsilon)\Gamma_R(\varepsilon)\big],
\end{equation}  
where $G^{r/a}_C$ is the retarded/advanced Green's
function of the central region. The coupling strength is given by
$\Gamma_{L/R}=i(\Sigma_{L/R}-\Sigma_{L/R}^{\dagger})$, where
$\Sigma_{L/R}$ is the self-energy due to the left/right lead. In 
the TB model the central region consists of the (infinite) array of
molecular states. 

We label all TB orbitals in the model, including the molecular states, by integer coordinates
$|n,m,l\rangle$ where $n$ refers to the direction of transport
($x$-axis) and $m,l$ refer to the perpendicular directions ($y,z$-axes). We say that an orbital $|n,m,l\rangle$ belongs to the
central supercell if $m,l\in\{1,\ldots,N\}$. 
For each orbital in the central supercell we form the Bloch states
\begin{equation}
|k_y,k_z;n,m,l\rangle=\sum_{M,L\in \mathbb{Z}}e^{i(k_yM+k_zL)}|n,m+MN,l+LN\rangle,
\end{equation}
where $k_y,k_z\in ]-\pi;\pi]$. Due to the periodicity of the system in the
directions perpendicular to the transport, all
matrices entering Eq.~(\ref{eq.trans}) are diagonal with respect to $\bold k=(k_y,k_z)$ and
consequently the total transmission
decomposes into a sum of $\bold k$-dependent transmissions, 
\begin{equation}
T(\bold k;\varepsilon)=\text{Tr}\left[
G_{C}^{r}(\bold k;\varepsilon)
\Gamma_{L}(\bold k;\varepsilon)
G^{a}_{C}(\bold k;\varepsilon)
\Gamma_{R}(\bold k;\varepsilon)
                  \right].
\end{equation}
The transmission
\emph{per} supercell is then
evaluated as
\begin{equation}
T(\varepsilon)=\frac{1}{\Omega_{\text{BZ}}}\int_{\text{1.BZ}}T(\bold
k;\varepsilon)\text{d}\bold k, 
\end{equation}
where $\Omega_{\text{BZ}}$ is the area
of the first Brillouin zone. In practice the integral is converted
into a finite sum $\int \text{d}\bold k/\Omega_{\text{BZ}}\to
\sum_{\bold k}W_{\bold k}$, where $W_{\bold k}$ represent the weights
of the discrete $\bold k$-points.

In order to distinguish between interference effects due to the
repeated supercells and electronic structure effects resulting from an
insufficient $\bold k$-point sampling we study the TB model in three
different limits which we denote by A, B and C. Limit A is the ideal situation where the result for the
transmission function has been converged both with respect to the
transverse dimensions of the supercell ($N$) and the denseness of the
$\bold k$-point sampling. This is the situation where no interference
between the repeated molecular states occur and it thus corresponds to
transmission through a single state. In B the transverse dimensions of the supercell
are fixed to $N=3$ and the $\bold k$-point
sampling is converged to account properly for the electronic structure. Limit
C is like B except that the 1.BZ is sampled only at the $\Gamma$-point.
\begin{figure}[!b]
\includegraphics[width=0.85\linewidth,angle=270]{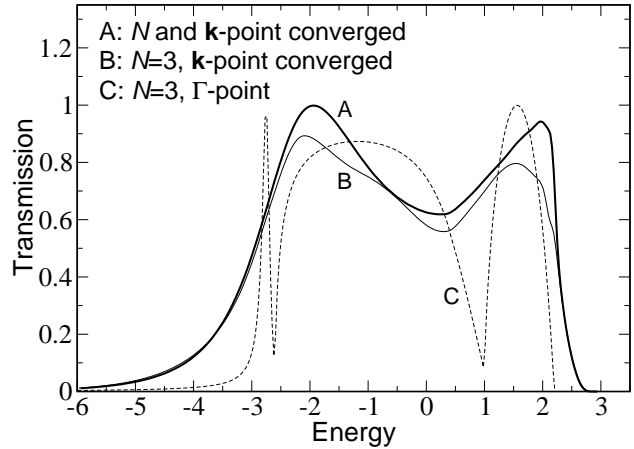}
\caption[cap.wavefct]{\label{fig2} Transmission through an array of
  molecular states in the $3d$ TB model with $\varepsilon_a=0$ and $t=2$. For A the result has been converged
  both with respect to $N$ and the number of $\bold k$-points. For B the
  number of $\bold k$-points has been converged, while C represents
  the case of $\Gamma$-point only. The difference
  between A and B is thus pure interference while B and C differ
  due to an insufficient $\bold k$-point sampling.}
\end{figure}

\begin{figure}[!b]
\includegraphics[width=0.95\linewidth]{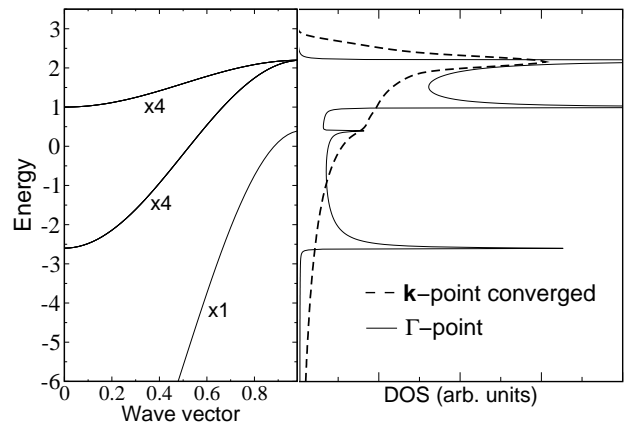}
\caption[cap.wavefct]{\label{fig3b} Left: Band diagram for 
the cubic TB lattice along the line $(k_x,0,0)$ corresponding to the
transport direction. The unit cell
contains $3 \times 3$ sites in the transverse plane. The
degeneracy of each band is indicated. Right: DOS
corresponding to the band diagram (full line) and the total DOS of the
TB lattice (dashed line) calculated
using a converged set of $\bold k$-points.  }
\end{figure}

In Fig.~\ref{fig2} we show the transmission function in each of the three
limits for the TB
model with parameters $\varepsilon_a=0$ and $t=2$. This choice for the
parameter $t$ corresponds to
rather strong coupling between the molecular states and the leads, and is
relevant for e.g. atomic metal contacts and chains or molecular
contacts in which the molecule forms a covalent bond with the leads.
The fully converged limit A is reached for $N=10$
and with $25$ $\bold k$-points, where all interference and $\bold k$-point
sampling effects have disappeared. The similarity between curves A
and B shows that interference effects are present but not very strong, even for
this small supercell with $N=3$. On the other hand a correct
description of the electronic
structure is crucial as can be seen from the large
discrepancy between curves A and C.

In order to understand the origin of the large difference between curves B and C,
we have calculated the relevant lead DOS for both cases. We use
a unit cell containing $3\times 3$ sites in the transverse plane and
calculate the DOS of the TB lattice using respectively the
$\Gamma$-point and a converged set of $\bold k$-points. The result is shown in the right
panel of Fig.~\ref{fig3b}. The $\Gamma$-point DOS has much more
structure than the total DOS, and exhibits pronounced peaks due to van Hove
singularities at the $1d$ band edges, see left panel. The strong van Hove
singularity is a characteristic of $1d$ systems and does generally not occur
in extended bulk systems. By sampling the 1.BZ in the transverse
plane, the density of van Hove singularities is increased and this
provides an effective smearing of the DOS.
By comparing the $\Gamma$-point DOS with curve C of Fig.~\ref{fig2} we
see that the dramatic dips in the transmission at energies $-2.6$ and $1.0$
coincide with two van Hove singularities. This clearly indicates that
the strong variation in the $\Gamma$-point transmission
is due to the artificial one dimensional description of the electronic structure of the lead. 

\begin{figure}[!b]
\includegraphics[width=0.85\linewidth,angle=270]{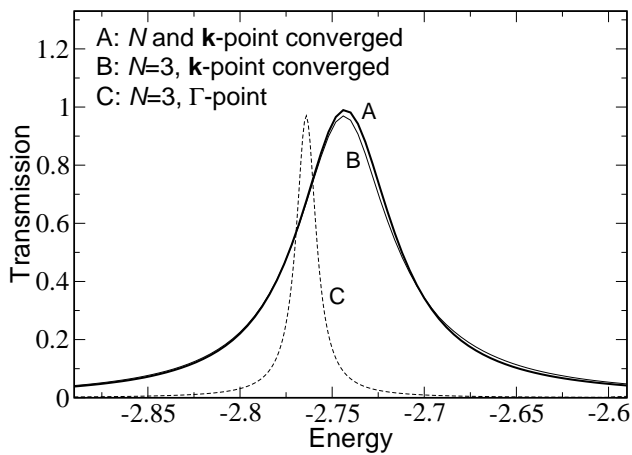}
\caption[cap.wavefct]{\label{fig4} Like Fig.~\ref{fig2}, except
  that the parameters $\varepsilon_a=-2.7$ and $t=0.3$ have been used.}
\end{figure}

As the coupling between the molecular states and the leads is
reduced, the transmission function narrows down and eventually turns
into a Lorentzian shaped resonance located close to the bare energy,
$\varepsilon_a$, with a width determined by the product of $t^2$ and the
DOS of the uncoupled lead at $\varepsilon_a$. This suggests that
the width of the resonance should be sensitive to the lead DOS, and in
particular that the $\Gamma$-point approximation could break down when
the resonance lies close to a van Hove singularity of the lead. That
this is in fact the case can be
seen in Fig.~\ref{fig4} which shows the transmission in the three
limits for $\varepsilon_a=-2.7$ and $t=0.3$. The curves A and B are
almost identical meaning that interference effects are 
negligible. In contrast the $\Gamma$-point
resonance is shifted too far down in energy and is about four times too
narrow as compared to the $\bold k$-point sampled result. We conclude that
$\bold k$-point sampling can be of importance also for weakly coupled
systems. We note, however, that the example shown in
Fig.~\ref{fig4} is a worst case
scenario due to the close proximity of $\varepsilon_a$ with a van Hove singularity.

\begin{figure}[!b]
\includegraphics[width=1.1\linewidth]{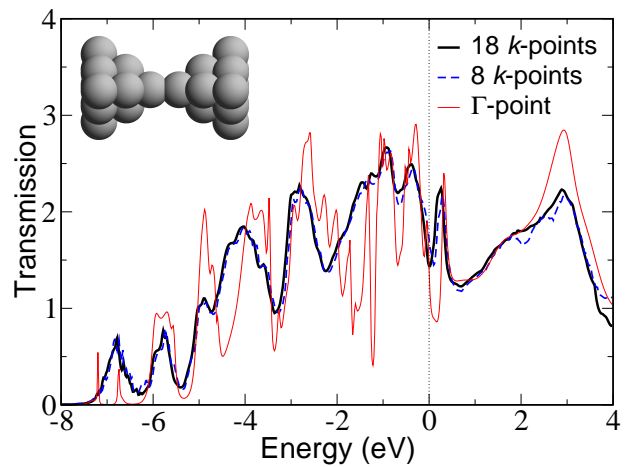}
\caption[cap.wavefct]{\label{fig5} (Color online) Transmission through the 
  Pt contact shown in the inset. The transmission has been calculated
  using 1, 8 and 18 $\bold k$-points in the transverse BZ.}
\end{figure}

As a second example we have calculated the transmission function for a Pt contact
using first principles techniques to evaluate the Hamiltonian. More
details on the computational method can be found
in Ref.~\onlinecite{method_chemphys}. The supercell contains $3\times
3$ atoms in the directions perpendicular to the transport direction
as sketched in the inset of Fig.~\ref{fig5}.
The transmission has been calculated using 1, 8 and 18
irreducible $\bold k$-points, and from the result we conclude that the
transmission function has converged to within $5\%$ using 8 $\bold
k$-points. In fact the same conclusion holds for several other systems
that we have studied, indicating that
convergence of the transmission function can in general be achieved using 8 irreducible $\bold
k$-points for a system containing $3\time 3$ atoms in the transverse cell.
Returning to Fig.~\ref{fig5} it can be seen that the $\Gamma$-point transmission has too much
structure and represents a rather poor approximation. This is very similar
to what we found for the TB model, except that the number of
unphysical dips/peaks in the transmission function is larger for the Pt contact. This is because
the effective number of orbitals in a cross-section of the supercell, which determines
the number of $1d$
bands in the lead and thus the density of van Hove
singularities, is larger in the first principles calculation than in the TB model.

Generally, the $\Gamma$-point approximation is expected to be valid
when the distance between the van Hove singularities is comparable
to the width of the features induced in the transmission function by
the van Hove singularities. It is our experience, both from TB
studies and first principles
calculations, that the transmission calculated at a single $\bold k$-point is
improved if the $\bold k$-point is chosen away from the
$\Gamma$-point. This is due to the lower degree of degeneracy among
the $1d$ lead bands at a general $\bold k$-point, which spreads out the van Hove singularities.

In conclusion, we have investigated the importance of 
$\bold k$-point sampling as well as 
interference among the repeated images in the supercell approach to coherent
electron transport. Using a representative TB model, it was found that 
the $\bold k$-point sampling is the most important factor of the two,
and that an insufficient $\bold k$-point sampling or, in particular, 
a $\Gamma$-point only sampling, can introduce
spurious features in the transmission function which could be wrongly
attributed to physical properties of the contact under study. The
origin of these spurious features was traced to the presence of van Hove
singularities in the lead.

We acknowledge support
from the Danish Center for Scientific Computing through Grant No.
HDW-1101-05.

%%%%%%% References

\bibliographystyle{apsrev}

\end{document}